\documentclass[preprint, superscriptaddress, showpacs,preprintnumbers,amsmath,amssymb]{revtex4}
\usepackage{graphicx}

\begin{document}

\thispagestyle{empty}

\title{Two approaches for describing the Casimir interaction
with graphene: density-density correlation function versus
polarization tensor}

\author{
G.~L.~Klimchitskaya}
\affiliation{Central Astronomical Observatory
at Pulkovo of the Russian Academy of Sciences,
St.Petersburg, 196140, Russia}
\affiliation{Institute of Physics, Nanotechnology and
Telecommunications, St.Petersburg State
Polytechnical University, St.Petersburg, 195251, Russia}
\author{
 V.~M.~Mostepanenko}
\affiliation{Central Astronomical Observatory
at Pulkovo of the Russian Academy of Sciences,
St.Petersburg, 196140, Russia}
\affiliation{Institute of Physics, Nanotechnology and
Telecommunications, St.Petersburg State
Polytechnical University, St.Petersburg, 195251, Russia}
\author{Bo~E.~Sernelius}
\affiliation{Division of Theory and Modelling, Department of Physics,
Chemistry and Biology, Link\"{o}ping University,
Link\"{o}ping, SE-581\,83, Sweden}

\begin{abstract}
The comparison studies of theoretical approaches to the
description of the Casimir interaction in layered systems
including graphene is performed. It is shown that at zero
temperature the approach using the polarization tensor
leads to the same results as the approach using the
longitudinal density-density correlation function of
graphene. An explicit expression for the zero-temperature
transverse density-density correlation function of
graphene is provided. We further show that the computational
results for the Casimir free energy of graphene-graphene and
graphene-Au plate interactions at room temperature, obtained
using the temperature-dependent polarization tensor,
deviate significantly from those using the longitudinal
density-density correlation function defined at zero temperature.
We derive both the longitudinal and transverse density-density
correlation functions of graphene at nonzero temperature.
The Casimir free energy in layered structures including
graphene, computed using the temperature-dependent correlation
functions, is exactly equal to that found using the polarization
tensor.
\end{abstract}

\maketitle
\section{Introduction}

During the last few years graphene and other carbon-based
nanostructures have attracted the particular attention of many
experimentallists and theorists due to their remarkable
properties \cite{1,2}. These investigations have provided
further impetus to technological progress.
One of the topical subjects, which came to the experimental
attention very recently \cite{3}, is the van der Waals and
Casimir interaction of graphene deposited on a substrate
with the test body made of an ordinary material.

Theorists have already undertook a number of studies of
graphene-graphene and graphene-material plate interactions
using the Dirac model of graphene \cite{1}, which assumes the
linear dispersion relation for the graphene bands at low
energies. Specifically, in Ref.~\cite{4} the van der Waals
coefficient for two graphene sheets at zero temperature was
calculated using the correlation energy from the random phase
approximation (in Ref.~\cite{5} the obtained value was improved
using the nonlocal dielectric function of graphene).
In Ref.~\cite{6}, the van der Waals and Casimir forces between
graphene and ideal metal plane were calculated at zero
temperature using the Lifshitz theory, where the reflection
coefficients of the electromagnetic oscillations were expressed
via the polarization tensor in (2+1)-dimensions.
The important progress was achieved in Ref.~\cite{7}, where
the force at nonzero temperature between two graphene sheets and
between a graphene and a material plate was expressed via the
Coulomb coupling between density fluctuations.
The density-density correlation function in the random-phase
approximation has been used. It was shown \cite{7} that for
graphene the relativistic effects are not essential, and that
the thermal effects become crucial at much shorter separations
than in the case of ordinary materials.
In Ref.~\cite{8}, the graphene-graphene interaction was computed
under an assumption that the conductivity of graphene can be
described by the in-plane optical properties of graphite.
It was shown \cite{8a} that for sufficiently large band gap
parameter of graphene the thermal Casimir force can vary
several-fold with temperature.
In Ref.~\cite{9}, the reflection coefficients in the Lifshitz
theory were expressed via the polarization tensor at nonzero
temperature whose components were explicitly calculated.
The detailed computations of graphene-graphene and
graphene-real metal Casimir interactions using this method
were performed \cite{9,10,11}. Finally, in Ref.~\cite{12}
the reflection coefficients of the Lifshitz theory were
generalized for the case of planar structures including
two-dimensional sheets. The graphene-graphene and
graphene-real metal interactions at both zero and nonzero
temperature were computed by using the electric susceptibility
(polarizability) of graphene expressed via the density-density
correlation function. It was argued \cite{12} that the
zero-temperature form of polarizability can be used also at
room temperature.

We underline that there is no complete agreement between the
results of different papers devoted to the van der Waals and
Casimir interactions with graphene (see Ref.~\cite{11} where
some of the results obtained are compared).
In fact, all the approaches go back to the Lifshitz theory
\cite{13,14,15}, but with different approximations made and
with various forms of the reflection coefficients used.
By and large the approaches based on the density-density
correlation function used its longitudinal version, i.e.,
neglected by the role of (small \cite{7}) relativistic effects.
Furthermore, dependence of the correlation function on
temperature which was unknown until the present time, was
obtained by means of scaling \cite{7} or even neglected
\cite{12}. By contrast, calculations based on the polarization
tensor are fully relativistic and include an explicit dependence
of its components on the temperature \cite{9,10,11}. This is
the reason why it would be useful to establish a link
between the two approaches and to test the validity of the
approximations used.

In this paper, we find a correspondence between the reflection
coefficients of the electromagnetic fluctuations on graphene
expressed in terms of electric susceptibility
(polarizability) of graphene  and components of the
polarization tensor. On this basis, we derive explicit
expressions for both longitudinal and transverse
electric susceptibilities of graphene,
density-density correlation function and
conductivities at arbitrary temperature.
Then we consider the limiting
cases of the obtained expressions at
zero temperature and find that the longitudinal version
coincides with that derived within the random phase
approximation. Furthermore, we compare the computational
results for graphene-graphene and graphene-real metal
interactions at room temperature obtained using the
polarization tensor \cite{10,11} with those obtained using
the density-density correlation function in Ref.~\cite{12}.
In doing so we pay special attention to contributions of the
transverse electric susceptibility of graphene and explicit
temperature dependence of the longitudinal
density-density correlation function to the Casimir
free energy.

The paper is organized as follows. In Sec.~II we establish
a link between the two approaches and derive the
density-density correlation functions at nonzero temperature.
Section~III is devoted to the case of zero temperature.
In Sec.~IV the computational results for graphene-graphene and
graphene-real metal thermal Casimir interactions using the
zero-temperature correlation function and the polarization tensor
at room temparature are compared. In Sec.~V the reader will find
our conclusions and discussion.

\section{Comparison between the reflection coefficients in two
theoretical approaches}

As discussed in Sec.~I, all theoretical approaches to the
van der Waals and Casimir interaction between two graphene sheets
or between graphene and material plate go back to the Lifshitz
theory representing the free energy per unit area
at temperature $T$ in thermal
equilibrium in the form \cite{13,14,15}
\begin{eqnarray}
&&
{\cal F}(a,T)=\frac{k_BT}{2\pi}\sum_{l=0}^{\infty}
{\vphantom{\sum}}^{\prime}\int_{0}^{\infty}\!\!
k_{\bot}dk_{\bot}\left\{\ln\left[1-r_{\rm TM}^{(1)}(i\xi_l,k_{\bot})
r_{\rm TM}^{(2)}(i\xi_l,k_{\bot})e^{-2aq_l}\right]
\right.
\nonumber \\
&&~~~~~
\left.
+\ln\left[1-r_{\rm TE}^{(1)}(i\xi_l,k_{\bot})
r_{\rm TE}^{(2)}(i\xi_l,k_{\bot})e^{-2aq_l}\right]\right\}.
\label{eq1}
\end{eqnarray}
\noindent
Here, $k_B$ is the Boltzmann constant, $k_{\bot}$ is the
projection of the wave vector on the plane of graphene,
$\xi_l=2\pi k_BTl/\hbar$ with $l=0,\,1,\,2,\,\ldots$ are
the Matsubara frequencies,
$q_l=(k_{\bot}^2+\xi_l^2/c^2)^{1/2}$, and the prime on the
summation sign indicates that the term with $l=0$ is
divided by two. The reflection coefficients on the two
boundary planes separated by the vacuum gap of width $a$
for the two independent polarizations of the electromagnetic
field, transverse magnetic (TM) and transverse electric (TE),
are notated as $r_{\rm TM,TE}^{(1)}$ and $r_{\rm TM,TE}^{(2)}$.

Let the first boundary plane be the freestanding graphene.
There are two main representations for the reflection
coefficents $r_{\rm TM,TE}^{(1)}\equiv r_{\rm TM,TE}^{(g)}$ on
graphene. We begin with the TM coefficient.
Within the first theoretical approach, the longitudinal
electric susceptibility (polarizability) of graphene at the
imaginary Matsubara frequencies is expressed as
\begin{equation}
\alpha^{||}(i\xi_l,k_{\bot})\equiv
\varepsilon^{||}(i\xi_l,k_{\bot})-1=
-\frac{2\pi e^2}{k_{\bot}}\chi^{||}(i\xi_l,k_{\bot}),
\label{eq2}
\end{equation}
\noindent
where $\chi^{||}(i\xi_l,k_{\bot})$ is the longitudinal
density-density correlation function. The latter is connected
with the dynamical conductivity of graphene by \cite{12}
\begin{equation}
\sigma^{||}(i\xi_l,k_{\bot})=
-\frac{e^2\xi_l}{k_{\bot}^2}\chi^{||}(i\xi_l,k_{\bot}),
\label{eq3}
\end{equation}
\noindent
where $e$ is the electron charge. Then the TM reflection
coefficient of the electromagnetic oscillations on graphene can
be expressed as \cite{12,16,17}
\begin{equation}
r_{\rm TM}^{(g)}(i\xi_l,k_{\bot})=
\frac{q_l\,\alpha^{||}(i\xi_l,k_{\bot})}{k_{\bot}+
q_l\,\alpha^{||}(i\xi_l,k_{\bot})}.
\label{eq4}
\end{equation}
\noindent
The explicit form for $\alpha^{||}$ is discussed below.

Within the second theoretical approach, the TM reflection
coefficient is expressed via the 00-component $\Pi_{00}$ of
the polarization tensor in (2+1)-dimensional space-time
\cite{9,10,11}
\begin{equation}
r_{\rm TM}^{(g)}(i\xi_l,k_{\bot})=
\frac{q_l\Pi_{00}(i\xi_l,k_{\bot})}{2\hbar k_{\bot}^2+
q_l\Pi_{00}(i\xi_l,k_{\bot})}.
\label{eq5}
\end{equation}
\noindent
The analytic expression for $\Pi_{00}$ is known \cite{9,10,11}.
It depends on the temperature both implicitly (through the
Matsubara frequencies) and explicitly, as on a parameter.
For the pristine (undoped) gapless graphene one has \cite{9,10,11}
\begin{eqnarray}
&&
\Pi_{00}(i\xi_l,k_{\bot})=
\frac{\pi\hbar\alpha k_{\bot}^2}{f(\xi_l,k_{\bot})}+
\frac{8\hbar\alpha c^2}{v_F^2}\int_{0}^{1}\!\!\!dx
\left\{
\vphantom{\frac{\xi_l^2\sqrt{x(1-x)}}{c^2f(\xi_l,k_{\bot})}}
\frac{k_BT}{\hbar c}\right.
\label{eq6}\\
&&~\times\ln\left[1+2\cos(2\pi lx)
e^{-\theta_T(\xi_l,k_{\bot},x)}+
e^{-2\theta_T(\xi_l,k_{\bot},x)}\right]
\nonumber \\
&&
~~~
-\frac{\xi_l}{2c}(1-2x)
\frac{\sin(2\pi lx)}{\cosh\theta_T(\xi_l,k_{\bot},x)+
\cos(2\pi lx)}
\nonumber \\
&&
~~~\left.
+\frac{\xi_l^2\sqrt{x(1-x)}}{c^2f(\xi_l,k_{\bot})}
\frac{\cos(2\pi lx)+e^{-\theta_T(\xi_l,k_{\bot},x)}}{\cosh\theta_T(\xi_l,k_{\bot},x)+
\cos(2\pi lx)}\right\},
\nonumber
\end{eqnarray}
\noindent
where $\alpha=e^2/(\hbar c)$ is the fine structure constant,
$v_F$ is the Fermi velocity, and the following notations are
introduced
\begin{eqnarray}
&&
f(\xi_l,k_{\bot})\equiv\left(\frac{v_F^2}{c^2}k_{\bot}^2+
\frac{\xi_l^2}{c^2}\right)^{1/2},
\label{eq7} \\
&&
\theta_T(\xi_l,k_{\bot},x)\equiv
\frac{\hbar c}{k_BT}f(\xi_l,k_{\bot})\sqrt{x(1-x)}.
\nonumber
\end{eqnarray}

Now we equate the right-hand sides of Eqs.~(\ref{eq4}) and
(\ref{eq5}) and obtain the expression for the longitudinal
polarizability of graphene at nonzero temperature via the
00-component of the polarization tensor
\begin{equation}
\alpha^{||}(i\xi_l,k_{\bot})=\frac{1}{2\hbar k_{\bot}}
\Pi_{00}(i\xi_l,k_{\bot}).
\label{eq8}
\end{equation}
\noindent
Using Eq.~(\ref{eq2}), for the longitudinal density-density
correlation function one obtains
\begin{equation}
\chi^{||}(i\xi_l,k_{\bot})=-\frac{1}{4\pi e^2\hbar}
\Pi_{00}(i\xi_l,k_{\bot}),
\label{eq9}
\end{equation}
\noindent
where $\Pi_{00}$ is given by Eq.~(\ref{eq6}). Similar to the
polarization tensor, the density-density correlation function
depends on $T$ both implicitly and explicitly.
The longitudinal conductivity of graphene at any $T$ is given
by Eq.~(\ref{eq3}).

We continue with the TE reflection coefficient.
Note that  Eqs.~(\ref{eq2}) and (\ref{eq3}) remain valid for
the transverse quantities:
the polarizability of graphene $\alpha^{\bot}(i\xi_l,k_{\bot})$,
the transverse permittivity $\varepsilon^{\bot}(i\xi_l,k_{\bot})$,
the density-density correlation function
$\chi^{\bot}(i\xi_l,k_{\bot})$,
and the conductivity $\sigma^{\bot}(i\xi_l,k_{\bot})$.
The TE reflection coefficient on graphene in terms of the
transverse polarizability was found in Ref.~\cite{12}
\begin{equation}
r_{\rm TE}^{(g)}(i\xi_l,k_{\bot})=
-\frac{\xi_l^2\alpha^{\bot}(i\xi_l,k_{\bot})}{c^2k_{\bot}q_l+
\xi_l^2\alpha^{\bot}(i\xi_l,k_{\bot})}.
\label{eq10}
\end{equation}
\noindent
Note that according to our knowledge no
explicit expression of $\alpha^{\bot}$ for graphene is available
in the published literature.

In terms of the polarization tensor, the TE reflection
coefficient takes the form \cite{9,10,11}
\begin{equation}
r_{\rm TE}^{(g)}(i\xi_l,k_{\bot})=-
\frac{k_{\bot}^2\Pi_{\rm tr}(i\xi_l,k_{\bot})-
q_l^2\Pi_{00}(i\xi_l,k_{\bot})}{2\hbar k_{\bot}^2q_l+
k_{\bot}^2\Pi_{\rm
tr}(i\xi_l,k_{\bot})-q_l^2\Pi_{00}(i\xi_l,k_{\bot})},
\label{eq11}
\end{equation}
\noindent
where the index tr denotes the sum of spatial component
$\Pi_1^{\,1}$ and $\Pi_2^{\,2}$.
For the undoped gapless graphene the analytic expression for
$\Pi_{\rm tr}$ is the following \cite{9,10,11}:
\begin{eqnarray}
&&
\Pi_{\rm tr}(i\xi_l,k_{\bot})=\Pi_{00}(i\xi_l,k_{\bot})+
\frac{\pi\hbar\alpha}{f(\xi_l,k_{\bot})}\left[f^2(\xi_l,k_{\bot})+
\frac{\xi_l^2}{c^2}\right]
\label{eq12}\\
&&
+8\hbar\alpha\int_{0}^{1}\!\!\!dx
\left\{
\vphantom{\frac{\xi_l^2\sqrt{x(1-x)}}{c^2f(\xi_l,k_{\bot})}}
\frac{\xi_l}{c}(1-2x)
\frac{\sin(2\pi lx)}{\cosh\theta_T(\xi_l,k_{\bot},x)+
\cos(2\pi lx)}\right.
\nonumber \\
&&
~~~\left.
-\frac{\sqrt{x(1-x)}}{f(\xi_l,k_{\bot})}\left[f^2(\xi_l,k_{\bot})+
\frac{\xi_l^2}{c^2}\right]
\frac{\cos(2\pi lx)+e^{-\theta_T(\xi_l,k_{\bot},x)}}{\cosh\theta_T(\xi_l,k_{\bot},x)+
\cos(2\pi lx)}\right\}.
\nonumber
\end{eqnarray}

By equating the right-hand sides of Eqs.~(\ref{eq10}) and
(\ref{eq11}), one obtains the expression for the transverse
polarizability of graphene at any nonzero temperature
\begin{equation}
\alpha^{\bot}(i\xi_l,k_{\bot})=
\frac{c^2}{2\hbar k_{\bot}\xi_l^2}\left[k_{\bot}^2
\Pi_{\rm tr}(i\xi_l,k_{\bot})-
q_l^2\Pi_{00}(i\xi_l,k_{\bot})\right].
\label{eq13}
\end{equation}
\noindent
The respective result for the transverse
density-density correlation function is found from equation
similar to Eq.~(\ref{eq2})
\begin{equation}
\chi^{\bot}(i\xi_l,k_{\bot})=
-\frac{c^2}{4\pi\hbar e^2\xi_l^2}\left[k_{\bot}^2
\Pi_{\rm tr}(i\xi_l,k_{\bot})-
q_l^2\Pi_{00}(i\xi_l,k_{\bot})\right].
\label{eq14}
\end{equation}
\noindent
Then the transverse conductivity of graphene is given by
Eq.~(\ref{eq3}) where the index $||$ is replaced with $\bot$.

We emphasize that Eqs.~(\ref{eq4}), (\ref{eq5}) and (\ref{eq10}),
(\ref{eq11}) are the exact consequencies of the Maxwell equations
and electrodynamic boundary conditions imposed on the 2D graphene
sheet. For this reason, the obtained connections (\ref{eq8}),
(\ref{eq9}) and (\ref{eq13}), (\ref{eq14}) between the
polarizabilities and
density-density correlation functions for graphene, on the one
hand, and the components of the polarization tensor, on the other
hand, are the exact ones. Keeping in mind that Eqs.~(\ref{eq6})
and (\ref{eq12}) for the polarization tensor  are calculated in
the one-loop approximation \cite{9}, the specific expressions
for the polarizabilities and
density-density correlation functions obtained after the
substitution of Eqs.~(\ref{eq6}) and (\ref{eq12}) in
Eqs.~(\ref{eq8}), (\ref{eq9})
and (\ref{eq13}), (\ref{eq14}) should be also considered as
found in the same approximation. In the next section we compare
them with those contained in the literature.

To conclude this section, we present an explicit expression
for the quantity $k_{\bot}^2\Pi_{\rm tr}-q_l^2\Pi_{00}$
entering the transverse polarizability, the  density-density
correlation function and the conductivity of graphene.
Substituting $\Pi_{00}$ from Eq.~(\ref{eq6}) and
$\Pi_{\rm tr}$ from Eq.~(\ref{eq12}), one obtains after
identical transformations
\begin{eqnarray}
&&
k_{\bot}^2\Pi_{\rm tr}(i\xi_l,k_{\bot})-
q_l^2\Pi_{00}(i\xi_l,k_{\bot})=
\pi\hbar\alpha k_{\bot}^2f(\xi_l,k_{\bot})
\label{eq15}\\
&&~
-\frac{8\hbar\alpha c^2}{v_F^2}\int_{0}^{1}\!\!\!dx
\left\{
\vphantom{\frac{\xi_l^2\sqrt{x(1-x)}}{c^2f(\xi_l,k_{\bot})}}
\frac{k_BT\xi_l^2}{\hbar c^3}
\ln\left[1+2\cos(2\pi lx)
e^{-\theta_T(\xi_l,k_{\bot},x)}+
e^{-2\theta_T(\xi_l,k_{\bot},x)}\right]\right.
\nonumber \\
&&
~~~
-\left[2f^2(\xi_l,k_{\bot})-\frac{\xi_l^2}{c^2}\right]
\frac{\xi_l}{2c}(1-2x)
\frac{\sin(2\pi lx)}{\cosh\theta_T(\xi_l,k_{\bot},x)+
\cos(2\pi lx)}
\nonumber \\
&&
~~~\left.
+\sqrt{x(1-x)}
f^3(\xi_l,k_{\bot})
\frac{\cos(2\pi lx)+e^{-\theta_T(\xi_l,k_{\bot},x)}}{\cosh\theta_T(\xi_l,k_{\bot},x)+
\cos(2\pi lx)}\right\}.
\nonumber
\end{eqnarray}
\noindent
This expression is used in below calculations.

\section{Energy of the Casimir interaction between two graphene
sheets at zero temperature}

In the limiting case $T\to 0$ the summation over the discrete
Matsubara frequencies in Eq.~(\ref{eq1}) is replaced with
integration over
the imaginary frequency axis, and for two graphene sheets one
arrives to the Casimir energy per unit area
\begin{eqnarray}
&&
E(a,T)=\frac{\hbar}{4\pi^2}\int_{0}^{\infty}\!\!
k_{\bot}dk_{\bot}\int_{0}^{\infty}\!d\xi\left\{
\ln\left[1-{r_{\rm TM}^{(g)}}^2(i\xi,k_{\bot})
e^{-2aq}\right]
\right.
\nonumber \\
&&~~~~~
\left.
+\ln\left[1-{r_{\rm TE}^{(g)}}^2(i\xi,k_{\bot})
e^{-2aq}\right]\right\}.
\label{eq16}
\end{eqnarray}
\noindent
Here, the reflection coefficients are given by either
Eqs.~(\ref{eq4}) and (\ref{eq10}) or (\ref{eq5}) and
(\ref{eq11}), where the discrete frequencies $\xi_l$ are
replaced with the continuous $\xi$.

We begin from the contribution of the TM mode, $E_{\rm TM}$,
to the total energy (\ref{eq16}). In terms of the polarization
tensor, the reflection coefficient $r_{\rm TM}^{(g)}$ is
given by Eqs.~(\ref{eq5}) and (\ref{eq6}) with the notation
(\ref{eq7}). As can be seen in Eq.~(\ref{eq7}), the quantity
$\theta_T(\xi_l,k_{\bot},x)\to\infty$ when $T\to 0$.
Because of this, from Eq.~(\ref{eq6}) at $T=0\,$K one obtains
\cite{6}
\begin{equation}
\Pi_{00}(i\xi,k_{\bot})=
\frac{\pi\hbar\alpha k_{\bot}^2}{f(\xi,k_{\bot})}.
\label{eq17}
\end{equation}
\noindent
Using the notation (\ref{eq7}), one obtains from Eq.~(\ref{eq8})
the longitudinal polarizability of graphene at zero temperature
\begin{equation}
\alpha^{||}(i\xi,k_{\bot})=
\frac{\pi e^2}{2\hbar}\,
\frac{k_{\bot}}{\sqrt{v_F^2k_{\bot}^2+\xi^2}},
\label{eq18}
\end{equation}
\noindent
and from Eq.~(\ref{eq9}) the respective density-density
correlation function
\begin{equation}
\chi^{||}(i\xi,k_{\bot})=
-\frac{1}{4\hbar}\,
\frac{k_{\bot}^2}{\sqrt{v_F^2k_{\bot}^2+\xi^2}}.
\label{eq19}
\end{equation}
\noindent
The longitudinal conductivity of graphene at $T=0\,$K is obtained
from Eqs.~(\ref{eq3}) and (\ref{eq19}).

The density-density correlation function (\ref{eq19}) at
$T=0\,$K, derived from the polarization tensor, coincides with the
classical result \cite{18,19} which was used in computations of
Ref.~\cite{12}. Then, for the TM reflection coefficient on
graphene at $T=0\,$K we obtain one and the same result either from
 Eqs.~(\ref{eq4}) and (\ref{eq18}) or from  Eqs.~(\ref{eq5}) and
(\ref{eq17})
\begin{equation}
r_{\rm TM}^{(g)}(i\xi,k_{\bot})=
\frac{\pi e^2 \sqrt{c^2k_{\bot}^2+\xi^2}}{2\hbar c
\sqrt{v_F^2k_{\bot}^2+\xi^2}+\pi e^2 \sqrt{c^2k_{\bot}^2+\xi^2}}.
\label{eq20}
\end{equation}
\noindent
This reflection coefficient coincides with that used in
Ref.~\cite{12}.

We continue by considering the contribution of the TE mode,
$E_{\rm TE}$, to the Casimir energy (\ref{eq16}).
In terms of the polarization tensor, the reflection coefficient
$r_{\rm TE}^{(g)}$ is given by Eq.~(\ref{eq11}).
The combination of the components of the polarization tensor,
$k_{\bot}^2\Pi_{\rm tr}-q_l^2\Pi_{00}$, entering
Eq.~(\ref{eq11}), is given by Eq.~(\ref{eq15}). In the limiting
case $T\to 0$, one obtains from Eq.~(\ref{eq15})
\begin{equation}
k_{\bot}^2\Pi_{\rm tr}(i\xi,k_{\bot})-q^2\Pi_{00}(i\xi,k_{\bot})=
\pi\hbar\alpha k_{\bot}^2f(\xi,k_{\bot}).
\label{eq21}
\end{equation}
\noindent

Substituting Eq.~(\ref{eq21}) in Eq.~(\ref{eq13}), we find
the transverse polarizability of graphene at zero temperature
\begin{equation}
\alpha^{\bot}(i\xi,k_{\bot})=
\frac{\pi e^2k_{\bot}}{2\hbar\xi^2}\,
\sqrt{v_F^2k_{\bot}^2+\xi^2}.
\label{eq22}
\end{equation}
\noindent
In a similar way,
substituting Eq.~(\ref{eq21}) in Eq.~(\ref{eq14}), we find
the transverse  density-density correlation function
at $T=0\,$K
\begin{equation}
\chi^{\bot}(i\xi,k_{\bot})=
-\frac{k_{\bot}^2}{4\hbar\xi^2}\,
\sqrt{v_F^2k_{\bot}^2+\xi^2}.
\label{eq23}
\end{equation}

 The TE reflection coefficient at $T=0\,$K is obtained
  either substituting
 Eq.~(\ref{eq21}) in Eq.~(\ref{eq11}) or Eq.~(\ref{eq22}) in
Eq.~(\ref{eq10}). The result is
\begin{equation}
r_{\rm TE}^{(g)}(i\xi,k_{\bot})=
-\frac{\pi e^2 \sqrt{v_F^2k_{\bot}^2+\xi^2}}{2\hbar c
\sqrt{c^2k_{\bot}^2+\xi^2}+\pi e^2 \sqrt{v_F^2k_{\bot}^2+\xi^2}}.
\label{eq24}
\end{equation}
\noindent
As is seen from the comparison of Eqs.~(\ref{eq20}) and
(\ref{eq24}), the reflection coefficient $r_{\rm TE}^{(g)}$
has the opposite sign, as compared with $r_{\rm TM}^{(g)}$,
and its magnitude is obtained from the latter by the
interchanging of $c$ and $v_F$.

\begin{figure}[b]
\vspace*{-7cm}
\centerline{\hspace*{2cm}
\includegraphics{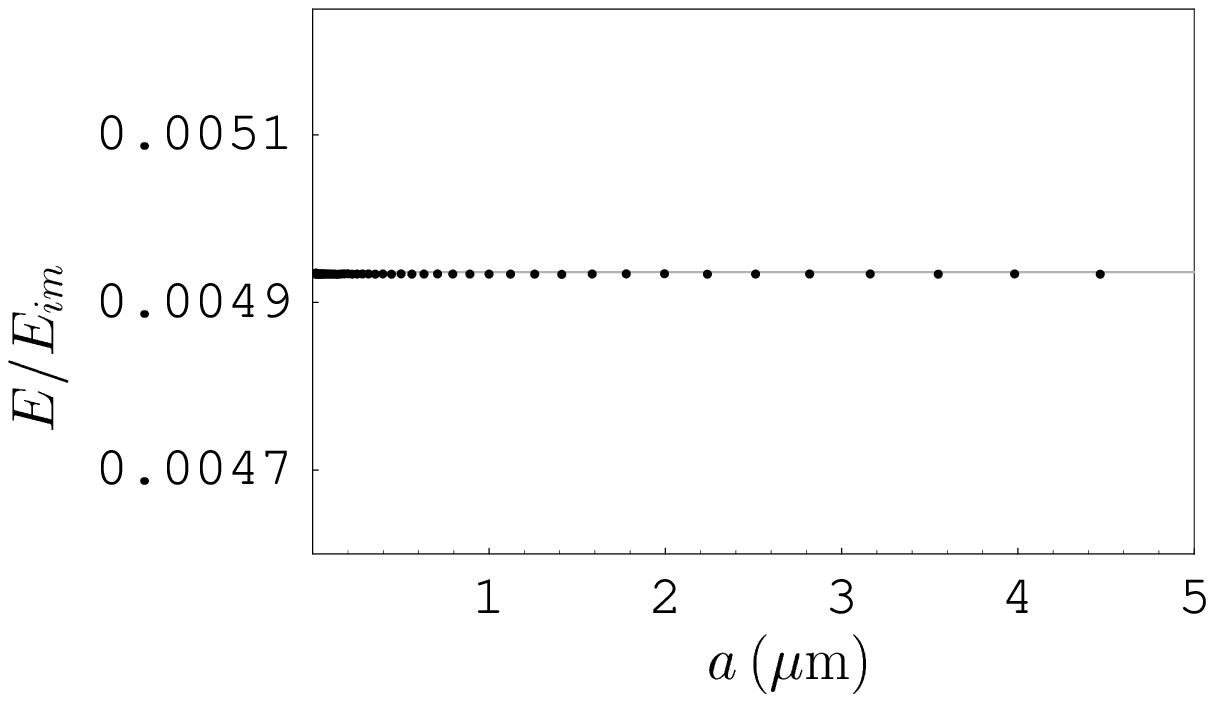}
}
\vspace*{-10cm}
\caption{\label{fg1}
The Casimir energy per unit area of two graphene sheets at
zero temperature normalized to that of two ideal-metal
planes is computed using the longitudinal density-density
correlation function (dots) and by the polarization tensor
(solid line), as functions of separation.
}
\end{figure}

Now we compare the computational results for the Casimir
energy per unit area of two parallel graphene sheets at zero
temperature obtained in Ref.~\cite{12} by means of the
density-density correlation function  and here using the
polarization tensor. In both cases the Fermi velocity
$v_F=8.73723\times 10^{5}\,$m/s is employed \cite{12,20,21}.
In  Fig.~\ref{fg1} the computational results of Ref.~\cite{12}
for $E(a)$ normalized for the Casimir energy per unit area of
two parallel ideal-metal planes
\begin{equation}
E_{im}(a)=-\frac{\pi^2}{720}\,\frac{\hbar c}{a^3}
\label{eq25}
\end{equation}
\noindent
are shown as black dots over the separation region from 10\,nm
to $5\,\mu$m. In making computations it was assumed \cite{12} that
$\chi^{\bot}(i\xi,k_{\bot})=\chi^{||}(i\xi,k_{\bot})$.
The gray line shows our computational results for
$E(a)/E_{im}(a)$ using the polarization tensor at $T=0\,$K given
by Eqs.~(\ref{eq17}) and (\ref{eq21}). In this case the
contribution of the TE mode was calculated precisely.

As can be seen in Fig.~\ref{fg1}, both sets of computational
results are in a very good agreement. This is explained by the
fact that $E_{\rm TM}(a)$ contributes 99.6\% of $E(a)$ and
$E_{\rm TE}(a)=0.004E(a)$ at all separation distances.
Furthermore, the relative differences between the
computational results of Ref.~\cite{12} for $E_{\rm TE}(a)$
(obtained under the assumption that $\chi^{\bot}=\chi^{||}$)
 and our results here
computed with the exact reflection coefficient $r_{\rm TE}^{(g)}$
are of about 0.1\%. Thus, the role of the TE contribution to the
Casimir energy of two graphene sheets is really negligibly
small \cite{7}, and it is  not critical what form of the
transverse
density-density correlation function is used in computations.
Physically this is connected with the fact that the TE contribution
is missing in the nonrelativistic limit, whereas the relativistic
effects contain additional small factors of the order of
$v_F/c$.

\section{Casimir interaction with graphene at nonzero temperature}

In this section we compare the computational results for the
Casimir free energy of two graphene sheets and a freestanding graphene
sheet interacting with an Au plate obtained using the approach of
Ref.~\cite{12} and using the polarization tensor. All computations
here are done at room temperature $T=300\,$K.
In this way we find the role of explicit dependence of the
density-density correlation function and polarization tensor on
the temperature.

\subsection{Two graphene sheets}

The free energy of the Casimir interaction between two sheets of
undoped graphene was computed at $T=300\,$K using Eq.~(\ref{eq1})
with
$r_{\rm TM,TE}^{(1)}=r_{\rm TM,TE}^{(2)}=r_{\rm TM,TE}^{(g)}$.
All computations were performed using the following two
approaches: the approach of Ref.~\cite{12} using the reflection
coefficients (\ref{eq4}) and (\ref{eq10}), expressed via the
zero-temperature longitunidal density-density correlation
function (\ref{eq19}) and the approach of
Ref.~\cite{11} using the reflection
coefficients (\ref{eq5}) and (\ref{eq11}) expressed via the
components of the polarization tensor (\ref{eq6}) and
(\ref{eq12}).
Within the approach Ref.~\cite{12}, the dependence of the free
energy on $T$ is determined by the $T$-dependent Matsubara
frequencies, whereas in the approach of Ref.~\cite{11} there
is also explicit dependence of the polarization tensor on $T$
as a parameter.

\begin{figure}[b]
\vspace*{-7cm}
\centerline{\hspace*{2cm}
\includegraphics{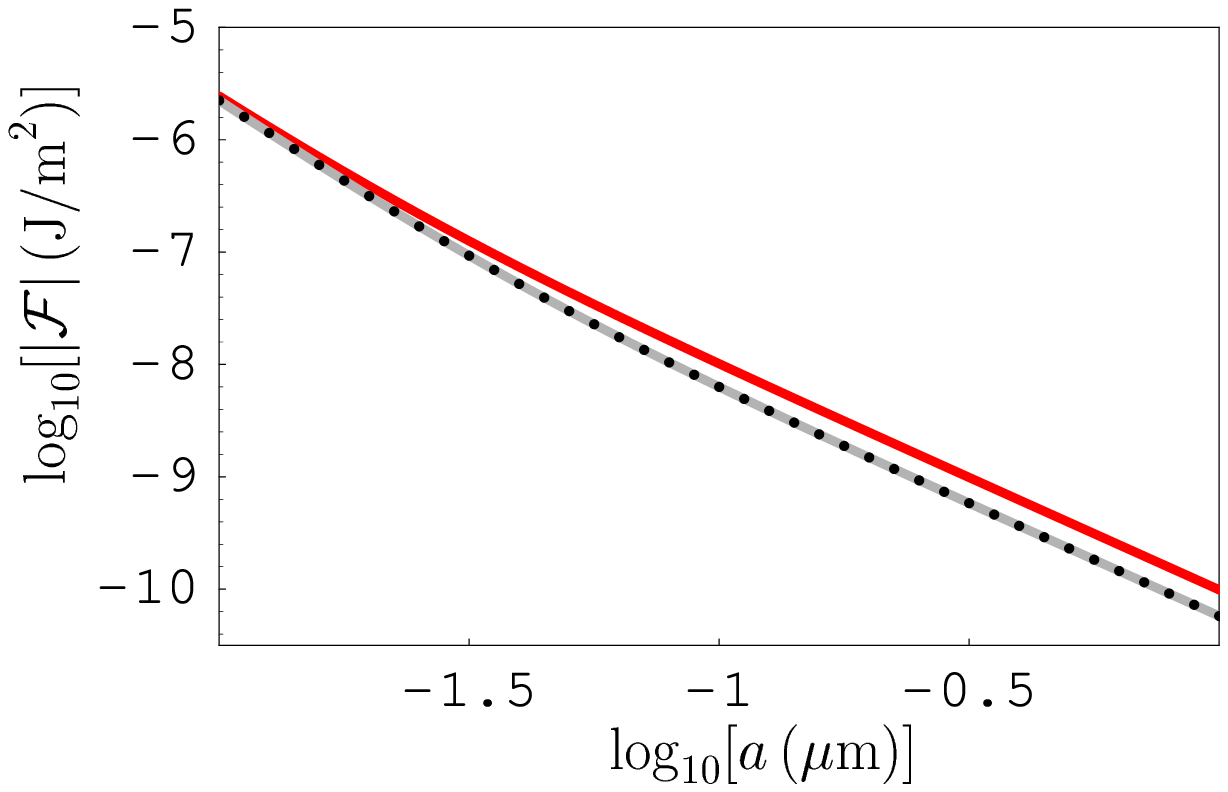}
}
\vspace*{-10cm}
\caption{\label{fg2}(Color online)
The magnitude of the Casimir free energy per unit area for
two graphene sheets at $T=300\,$K
is shown as a function of separation
in the logarithmic scale.
The upper and lower solid lines are computed using
the polarization tensor at $T=300\,$K and at $T=0\,$K,
respectively, whereas dots indicate the computational
results using the longitudinal
density-density correlation function
at $T=0\,$K.
}
\end{figure}

In Fig.~\ref{fg2} we present the computational results for the
Casimir free energy of two graphene sheets at $T=300\,$K
as functions of separation over the interval from 10\,nm
to $1\,\mu$m. The results obtained using the polarization tensor
at $T=300\,$K are shown as the upper solid line, and the
results obtained using the longitudinal density-density
correlation function (\ref{eq19}) defined at $T=0\,$K are shown
as dots.
{} From Fig.~\ref{fg2} it is seen that the upper solid line
deviates from dots significantly even at short separations.
This is explained by the dependence of the polarization tensor on
$T$ as a parameter in addition to the implicit $T$-dependence
through the Matsubara frequencies. The lower (gray) solid line
in Fig.~\ref{fg2} shows the computational results obtained by
means of the polarization tensor (\ref{eq17}) and (\ref{eq21})
at $T=0\,$K. This solid line is in a very good agreement with
dots computed using the formalism of Ref.~\cite{12}, as it
should be according to the results of Secs.~II and III.

Note that the dominant contribution to the free energy of
graphene-graphene interaction plotted in Fig.~\ref{fg2} is
given by the TM mode. Thus, at $a=10\,$nm
${\cal F}_{\rm TM}=0.9965{\cal F}$ and
${\cal F}_{\rm TE}=0.0035{\cal F}$.
Computations show that the contribution of the TM mode to the
total free energy increases with the increase of separation.
As a result, at $a=100\,$nm it holds
${\cal F}_{\rm TM}=0.9992{\cal F}$ and  at $a=1\,\mu$m
${\cal F}_{\rm TM}=0.9999{\cal F}$.

It should be stressed also that the deviation between the upper
and lower lines in Fig.~\ref{fg2} is explained entirely by the
thermal dependence of the polarization tensor at zero
Matsubara frequency. As was shown in Refs.~\cite{9,10,11}
(see also Ref.~\cite{22}) contributions of all Matsubara terms
with $l\geq 1$ are nearly the same irrespective of weather
the polarization tensor at $T=0\,$K or at $T\neq 0\,$K is used in
computations.

In this respect we remind that the contribution of the
zero-frequency term, ${\cal F}_{l=0}$, to the total free energy
of two graphene sheets ${\cal F}$ is increasing with the
increase of separation, for example,
${\cal F}_{l=0}=0.32{\cal F}$ at $a=10\,$nm,
${\cal F}_{l=0}=0.946{\cal F}$ at $a=100\,$nm, and
${\cal F}_{l=0}=0.9994{\cal F}$ at $a=1\,\mu$m.
The classical limit is already achieved at $a=400\,$nm where
${\cal F}_{l=0}=0.996{\cal F}$.
At $a\geq 400\,$nm the Casimir free energy shown by the upper
line in Fig.~\ref{fg2} is given to high accuracy by the
asymptotic expression \cite{9,11,22a}
\begin{equation}
{\cal F}(a,T)\approx -\frac{k_BT\zeta(3)}{16\pi a^2}
\left[1-\frac{1}{4\alpha \ln2}\left(
\frac{v_F}{c}\right)^2\frac{\hbar c}{ak_BT}\right],
\label{eq26}
\end{equation}
\noindent
where $\zeta(z)$ is the Riemann zeta function.

This should be compared with the asymptotic free energy
\begin{equation}
{\cal F}(a,T)\approx -\frac{k_BT}{16\pi a^2}
{\rm Li}_3\left({r_0^{(g)}}^2\right),
\label{eq27}
\end{equation}
\noindent
where ${\rm Li}_3(z)$ is the polylogarithm function.
Equation (\ref{eq27}) is obtained using the approach of
Ref.~\cite{12} where the TM reflection coefficient
at $\xi_0=0$ is defined by Eqs.~(\ref{eq4}) and (\ref{eq18})
\begin{equation}
r_{\rm TM}^{(g)}(0,k_{\bot})\equiv r_0^{(g)}=
\frac{\pi e^2}{2\hbar v_F+\pi e^2}.
\label{eq28}
\end{equation}
\noindent
The asymptotic expression (\ref{eq27}) is in a very good
agreement with the lower solid line (and dots) in Fig.~\ref{fg2}
at $a\geq 400\,$nm.
We also notice that the contribution of the TE mode at $l=0$
to the free energy, ${\cal F}_{{\rm TE},\,l=0}$, is negligibly
small, as compared with contribution of the TM mode at $l=0$
and with the total free energy:
$|{\cal F}_{{\rm TE},\,l=0}|<2\times 10^{-9}
|{\cal F}_{{\rm TM},\,l=0}|$ and
$|{\cal F}_{{\rm TE},\,l=0}|<1.2\times 10^{-9}|{\cal F}|$
over the entire region of separations.

\begin{figure}[b]
\vspace*{-4cm}
\centerline{\hspace*{2cm}
\includegraphics{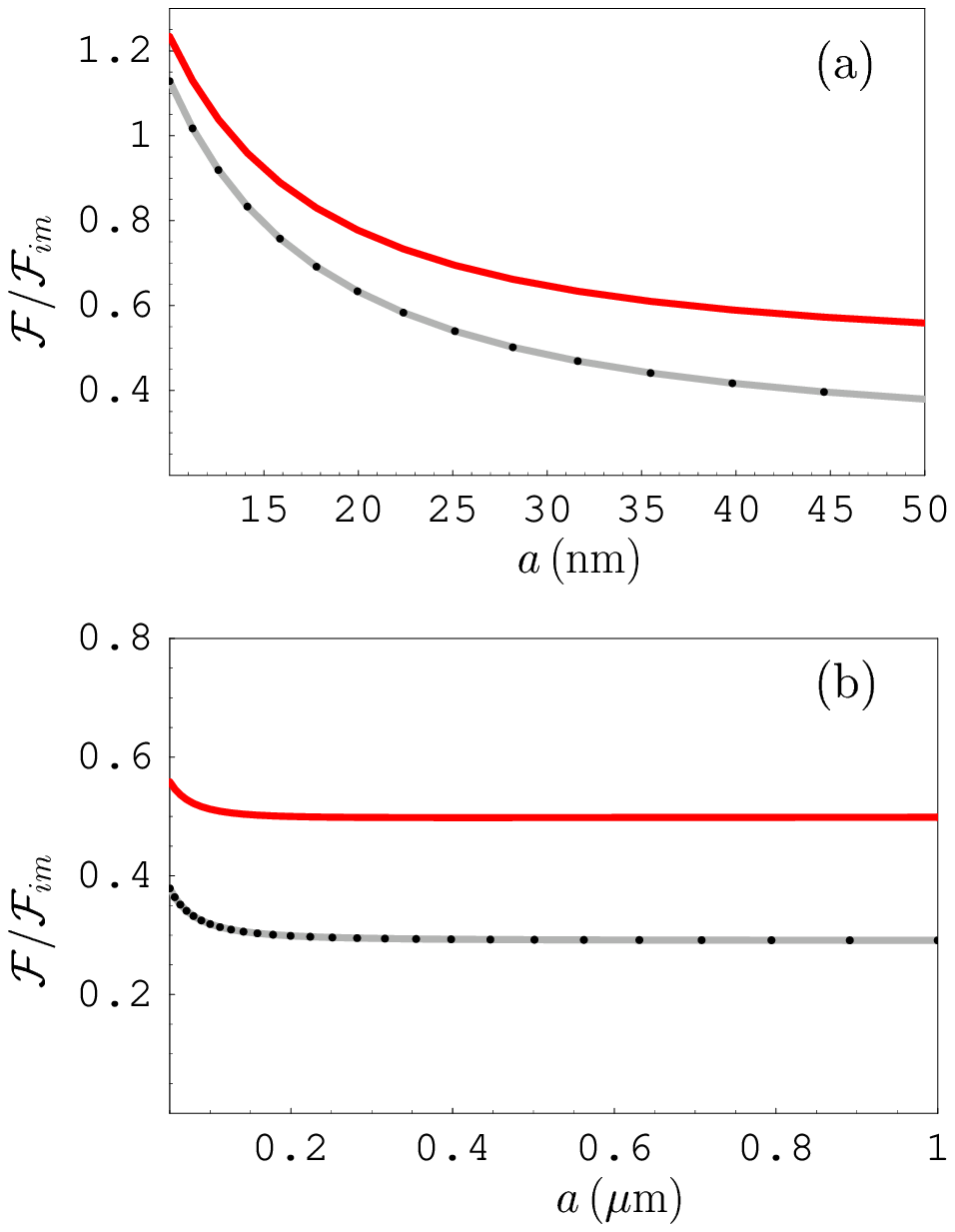}
}
\vspace*{-13cm}
\caption{\label{fg3}(Color online)
The Casimir free energy per unit area of
two graphene sheets at $T=300\,$K normalized to that
of two ideal-metal planes in the limit of high $T$ is
computed using the longitudinal density-density correlation
function at $T=0\,$K (dots), by the polarization tensor
at $T=300\,$K (the upper solid line) and
 by the polarization tensor at $T=0\,$K (the lower solid
 line) over the separation regions (a) from 10 to 50\,nm
 and (b) from 50\,nm to $1\,\mu$m.
}
\end{figure}

Now we present a more informative comparison between the
approaches using the polarization tensor at $T=300\,$K and
the longitudinal density-density correlation function at
$T=0\,$K, avoiding the use of the logarithmic scale.
For this purpose we plot in Fig.~\ref{fg3}(a,b)
the ratios of the obtained results for the free energy to
the asymptotic free energy of two ideal metal planes at
high temperature defined as \cite{15}
\begin{equation}
{\cal F}_{im}(a,T)=-\frac{k_BT\zeta(3)}{8\pi a^2}.
\label{eq29}
\end{equation}
\noindent
The upper and lower solid lines are computed by Eq.~(\ref{eq1})
using the polarization tensor at $T=300\,$K and $T=0\,$K,
respectively. The dots indicate the computational results of
Ref.~\cite{12} obtained at $T=300\,$K using the longitudinal
density-density correlation function defined at $T=0\,$K.
{}From Fig.~\ref{fg3}(a) it becomes clear that even at the
shortest separations from 10 to 50\,nm, where in the logarithmic
scale of Fig.~\ref{fg2} the computational results using the
two approaches might seem to be very close, there are in fact
large deviations illustrating the role of explicit thermal
dependence of the polarization tensor. In Fig.~\ref{fg3}(b)
plotted for the separation region from 50\,nm to $1\,\mu$m
it is seen that the high-temperature limits predicted by the
two approaches also differ significantly.
Note that the computational results shown by the upper
solid lines agree with those of Ref.~\cite{7} where the
temperature dependence of the longitudinal density-density
correlation function was found by scaling.

\begin{figure}[b]
\vspace*{-7cm}
\centerline{\hspace*{2cm}
\includegraphics{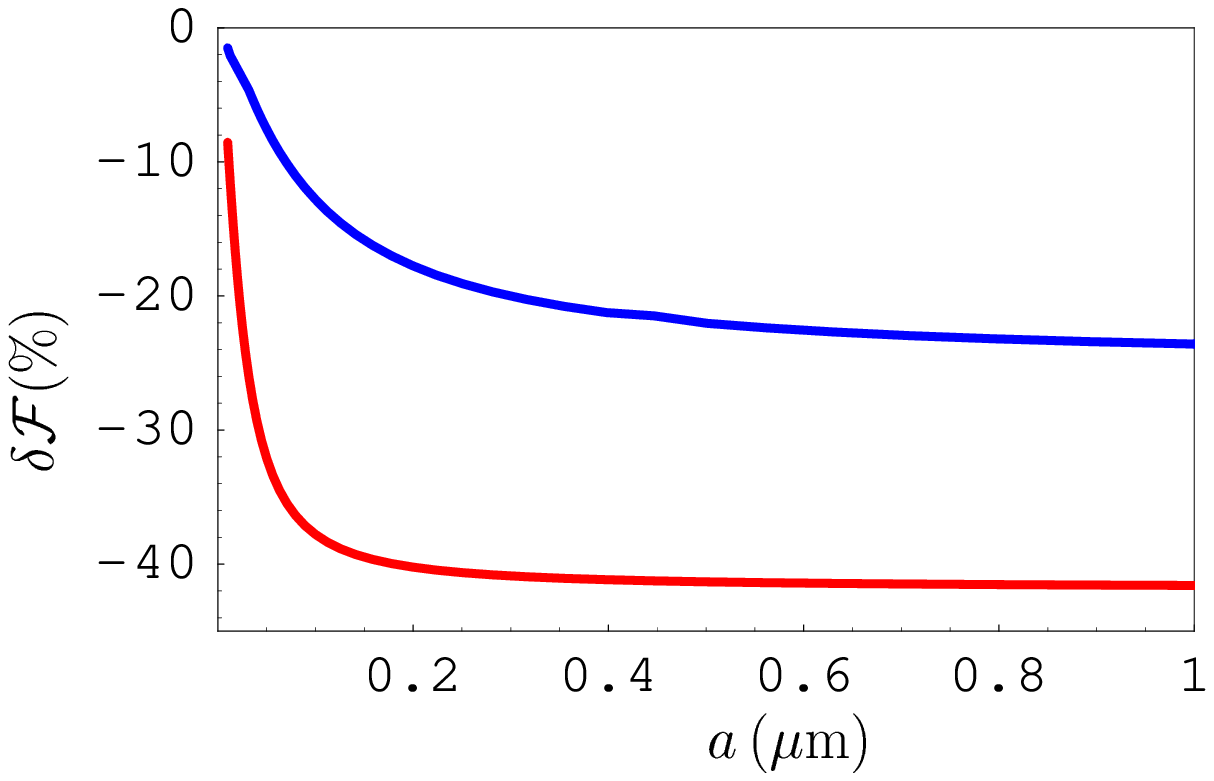}
}
\vspace*{-10cm}
\caption{\label{fg4}(Color online)
The solid lines show the relative deviations between the
Casimir free energies of graphene-graphene (the lower line)
and graphene-Au plate (the upper line) interactions computed using
the longitudinal density-density correlation
function at $T=0\,$K and the polarization tensor at $T=300\,$K.
}
\end{figure}

Finally, in Fig.~\ref{fg4} we plot by the lower solid line
the relative deviation between the free energies of two
graphene sheets computed using the longitudinal
density-density correlation function at $T=0\,$K
(${\cal F}_{dd}$) and the polarization tensor at $T=300\,$K
(${\cal F}_{pt}$)
\begin{equation}
\delta{\cal F}(a,T)=\frac{{\cal F}_{dd}(a,T)-
{\cal F}_{pt}(a,T)}{{\cal F}_{pt}(a,T)}.
\label{eq30}
\end{equation}
\noindent
As is seen in Fig.~\ref{fg4}, at the shortest separation
$a=10\,$nm the magnitude of the relative deviation
$|\delta{\cal F}|=8.5$\%, then it achieves the value of
$|\delta{\cal F}|=41.2$\% at $a=400\,$nm, and does not
exceed 41.8\% at all larger separations.

To conclude the consideration of two graphene sheets,
we stress that the calculation approach using the
temperature-dependent density-density correlation
functions (\ref{eq9}) and (\ref{eq14}) found in Sec.~II
and the reflection coefficients (\ref{eq4}) and (\ref{eq10})
lead to precisely the same results as the
temperature-dependent polarization tensor.

\subsection{Graphene sheet and a gold plate}

We have calculated the Casimir free energy at $T=300\,$K
for a graphene sheet interacting with an Au plate using two
theoretical approaches discussed above.
For this purpose Eq.~(\ref{eq1}) was used where the reflection
coefficients $r_{\rm TM,TE}^{(1)}=r_{\rm TM,TE}^{(g)}$ are
defined in Secs.~II and III and
$r_{\rm TM,TE}^{(2)}=r_{\rm TM,TE}^{(\rm Au)}$ are
defined as
\begin{eqnarray}
&&
r_{\rm TM}^{(\rm Au)}(i\xi_l,k_{\bot})=
\frac{\varepsilon(i\xi_l)q_l-k_l}{\varepsilon(i\xi_l)q_l
+k_l},
\nonumber \\
&&
r_{\rm TE}^{(\rm Au)}(i\xi_l,k_{\bot})=
\frac{q_l-k_l}{q_l+k_l}.
\label{eq31}
\end{eqnarray}
\noindent
Here, $\varepsilon(\omega)$ is the frequency-dependent dielectric
permittivity of Au and
\begin{equation}
k_l\equiv k_l(i\xi_l,k_{\bot})=\left[k_{\bot}^2+
\varepsilon(i\xi_l)\frac{\xi_l^2}{c^2}\right]^{1/2}.
\label{eq32}
\end{equation}

The dielectric permittivity of Au at the imaginary Matsubara
frequencies was obtained from the experimental optical
data \cite{23} for the imaginary part of the dielectric
function by means of the Kramers-Kronig relation.
The data were previously extrapolated to lower frequencies by
means of the Drude model. In this paper the data for
$\varepsilon(i\xi_l)$ from Ref.~\cite{12} have been used in
computations. The alternative extrapolation of the optical
data by means of the plasma model leads to a maximum relative
deviation in the obtained free energy equal to 0.8\% at the
shortest separation $a=10\,$nm and to smaller deviations at
larger separations.
As noted in Ref.~\cite{10}, for graphene-metal interaction
the Casimir free energy and pressure do not depend on what
model of metal (Drude or plasma) is used to describe the
metal. For two metallic plates there are large differences
in the results obtained using the Drude or plasma models
\cite{24} due to the contribution of the TE mode which is
negligibly small for a graphene sheet.

\begin{figure}[b]
\vspace*{-7cm}
\centerline{\hspace*{2cm}
\includegraphics{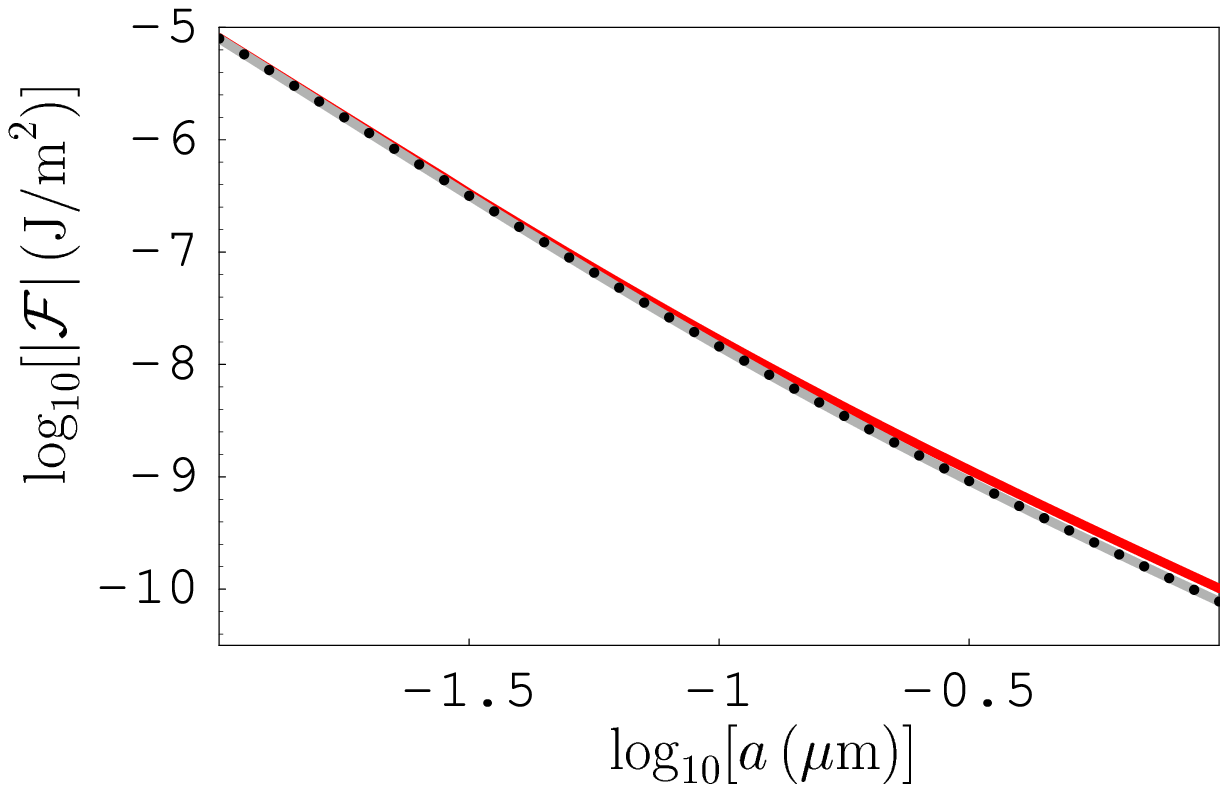}
}
\vspace*{-10cm}
\caption{\label{fg5}(Color online)
The magnitude of the Casimir free energy per unit area for
a graphene sheet and an Au plate at $T=300\,$K
 is shown as a function of separation
in the logarithmic scale.
The upper and lower solid lines are computed using
the polarization tensor at $T=300\,$K and at $T=0\,$K,
respectively, whereas dots indicate the computational
results using the longitudinal
density-density correlation function
at $T=0\,$K.
}
\end{figure}

In Fig.~\ref{fg5} the computational results for the Casimir
free energy of a graphene sheet interacting with an Au plate
at $T=300\,$K  are presented as functions of separation
in the region from 10\,nm to $1\,\mu$m. The upper and lower
solid lines indicate the results obtained using Eq.~(\ref{eq1})
and the polarization tensor at $T=300\,$K and $T=0\,$K,
respectively. The dots show the results \cite{12} computed
from the longitudinal density-density correlation function
(\ref{eq19}) at $T=0\,$K. As is seen in Fig.~\ref{fg5}, dots
are in agreement with the lower solid line, but deviate
significantly from the upper one. This demonstrates
important role of the explicit dependence of the polarization
tensor on the temperature.

As in the case of two graphene sheets, the dominant
contribution to ${\cal F}$ is given by the TM mode. Here,
however, the ratio ${\cal F}_{\rm TM}/{\cal F}$ is not a
monotonous function of $a$. Thus, at $a=10\,$nm
 ${\cal F}_{\rm TM}=0.983{\cal F}$ and at $a=100\,$nm
 the TM contribution achieves its minimum value
 ${\cal F}_{\rm TM}=0.961{\cal F}$.
 With further increase of separation  ${\cal F}_{\rm TM}$
increases to $0.983{\cal F}$, $0.992{\cal F}$, and
$0.998{\cal F}$ at $a=500\,$nm, $1\,\mu$m, and $2\,\mu$m,
respectively. Similar to the case of two graphene sheets,
the difference between the upper and lower solid lines is
explained by the explicit thermal dependence of the
polarization tensor at zero Matsubara frequency \cite{9,10}.

For a graphene sheet interacting with an Au plate, the
contribution of the zero Matsubara frequency to the total free
energy increases with separation slower than for two graphene
sheets. Thus, at $a=10\,$nm and 100\,nm one obtains
${\cal F}_{l=0}=0.11{\cal F}$ and $0.58{\cal F}$, respectively.
At $a=1\,\mu$m ${\cal F}_{l=0}=0.97{\cal F}$, and at
$a=1.6\,\mu$m the classical limit is achieved:
${\cal F}_{l=0}=0.99{\cal F}$. At this and larger separations
the Casimir free energy per unit area is given by \cite{9,11}
\begin{equation}
{\cal F}(a,T)\approx -\frac{k_BT\zeta(3)}{16\pi a^2}
\left[1-\frac{1}{8\alpha \ln2}\left(
\frac{v_F}{c}\right)^2\frac{\hbar c}{ak_BT}\right].
\label{eq33}
\end{equation}

The calculation approach using the longitudinal density-density
correlation function defined at zero temperature describes the
reflection coefficient on graphene at $\xi_0=0$ by
Eq.~(\ref{eq28}).
Taking into account that for Au
$r_{\rm TM}^{(\rm Au)}(0,k_{\bot})=1$, the classical limit is
obtained in the form
\begin{equation}
{\cal F}(a,T)\approx -\frac{k_BT}{16\pi a^2}
{\rm Li}_3\left({r_0^{(g)}}\right),
\label{eq34}
\end{equation}
\noindent
This asymptotic expression is in a very good agreement with
computational results of Ref.~\cite{12} at $a\geq 1.6\,\mu$m.

\begin{figure}[b]
\vspace*{-4cm}
\centerline{\hspace*{2cm}
\includegraphics{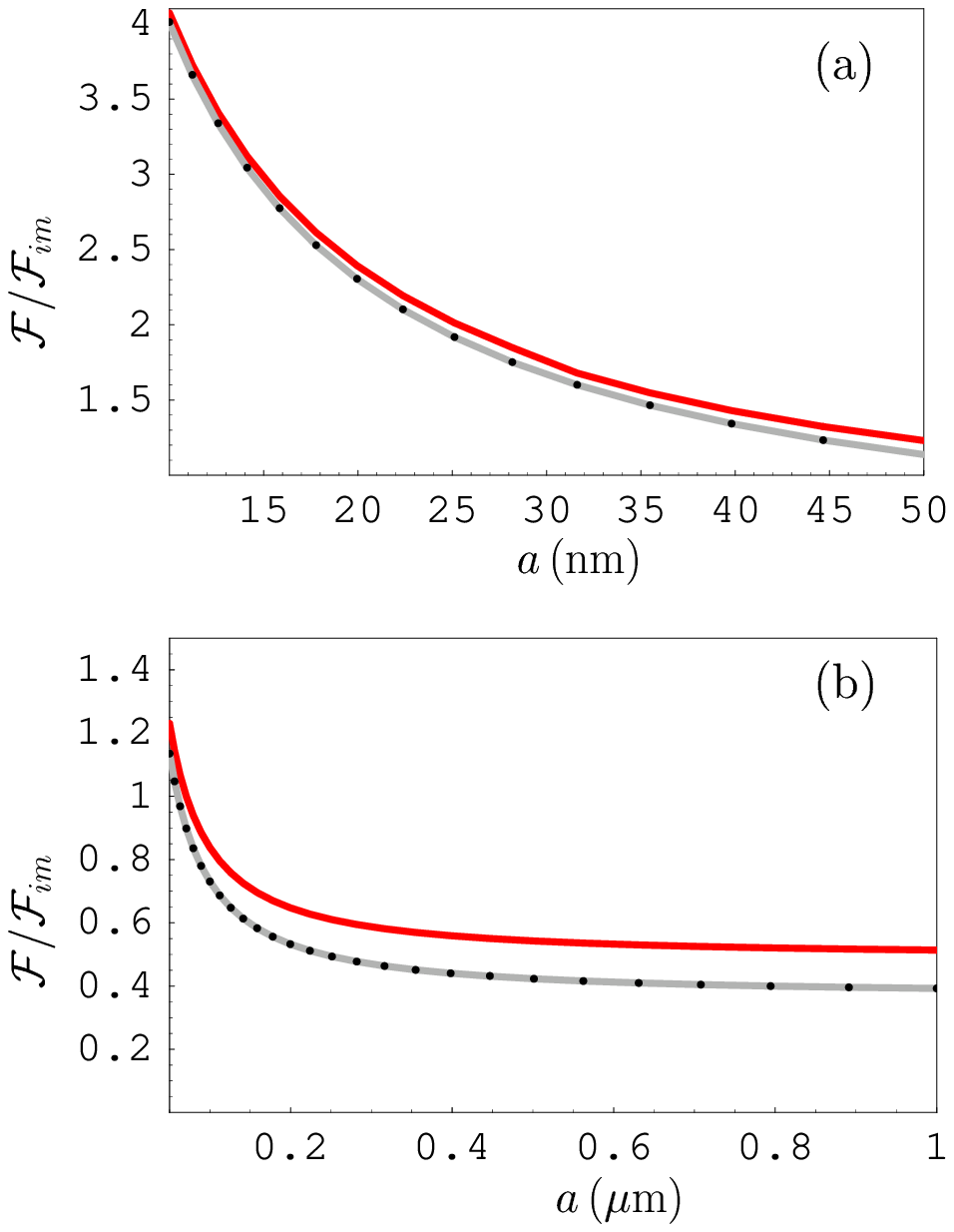}
}
\vspace*{-13cm}
\caption{\label{fg6}(Color online)
The Casimir free energy per unit area of
a graphene sheet and an Au plate
at $T=300\,$K normalized to that
of two ideal-metal planes in the limit of high $T$ is
computed using the longitudinal density-density correlation
function at $T=0\,$K (dots), by the polarization tensor
at $T=300\,$K (the upper solid line) and
 by the polarization tensor at $T=0\,$K (the lower solid
 line) over the separation regions (a) from 10 to 50\,nm
 and (b) from 50\,nm to $1\,\mu$m.
}
\end{figure}

To avoid the use of the logarithmic scale, in Fig.~\ref{fg6}(a,b)
we plot the ratios of the computed free energies of graphene-Au
plate interaction to the asymptotic free energy of two ideal-metal
planes (\ref{eq29}).
 The upper and lower
solid lines are computed by Eq.~(\ref{eq1}) using
the polarization tensor defined at $T=300\,$K and $T=0\,$K,
respectively. The dots are computed in Ref.~\cite{12}
at $T=300\,$K using
the longitudinal density-density correlation function
defined at $T=0\,$K. {}From  Figs.~\ref{fg6}(a) and \ref{fg6}(b)
it is seen that there are significant deviations between the
upper line, on the one hand, and the lower line and dots,
on the other hand, at both short and relatively large
separations. At $a\geq 1.6\,\mu$m the descrepancy between the
theoretical predictions of the two approaches is illustrated
by Eqs.~(\ref{eq33}) and (\ref{eq34}).

The relative deviation (\ref{eq30}) between the Casimir
free energies of graphene-Au plate interaction computed using the
two approaches is shown by the upper line in Fig.~\ref{fg4}.
Here, the magnitude of the relative deviation is equal to
$|\delta{\cal F}|=1.5$\% at $a=10\,$nm , achieves
$|\delta{\cal F}|=24$\% at $a=1.5\,\mu$m, and does not
exceed 24.5\% at larger separations.
This means that large thermal effects inherent to graphene
are less pronounced in the graphene-Au plate configuration, as
compared to the case of two graphene sheets.

Similar to two graphene sheets, for a graphene interaction
with an Au plate the theoretical predictions using the
polarization tensor at $T\neq 0$ are in full agreement with
respective predictions using the $T$-dependent density-density
correlation function defind in Eqs.~(\ref{eq9}) and (\ref{eq14}).

\section{Conclusions and discussion}

In the foregoing, we have performed the comparison studies
of two approaches used to calculate the van der Waals and
Casimir interaction between two graphene sheets and between
a graphene sheet and a metal plate.
One of these approaches is based on the use of the polarization
tensor. All its components are found \cite{9} at any temperature.
The other approach is based on the use of the  density-density
correlation function. Only the longitudinal version of this
function was available in the literature and only at zero
temperature. Because of this, previous calculations estimated the
TE contribution to the free energy as negligibly small and
either modeled the temperature dependence of the correlation
function by means of scaling between two asymptotic regimes
\cite{7} or argued that this dependence is not
essential \cite{12}.

We have shown that at zero temperature the approaches using the
polarization tensor and the standard longitudinal density-density
correlation function lead to almost coinciding computational
results. The coincidence becomes exact if in the calculation of
negligibly small contribution of the TE mode one replaces the
longitudinal density-density correlation function at $T=0\,$K
for the transverse one. We have provided an explicit expression
for this function.

Computations at nonzero temperature using the polarization tensor
with an explicit thermal dependence demonstrate significant
deviations
from the computational results using the
density-density correlation function at $T=0\,$K. The latter
include only an implicit dependence of the Casimir free energy
on the temperature through the Matsubara frequencies.
It was shown that for graphene-graphene and graphene-Au plate
interactions the free energies obtained using this approach
deviate from those calculated using the temperature-dependent
polarization tensor up to 41.8\% and 24.5\%, respectively.
However, the computational results obtained
using the zero-temperature
density-density correlation function are reproduced when the
polarization tensor defined at $T=0\,$K is used.
Similar to the case of zero temperature, at $T\neq 0\,$K the
contribution of the TE mode of the electromagnetic field to the
Casimir free energy is shown to be negligibly small for both
graphene-graphene and graphene-Au plate systems.

We have performed a comparison between the exact TM and TE
reflection coefficients expressed via the components of the
polarization tensor, on the one hand, and via the
longitudinal and transverse density-density correlation functions,
on the other hand. In this way we have found explicit
expressions for both longitudinal and transverse density-density
correlation functions at any nonzero temperature.
In the limiting case of vanishing temperature, our
temperature-dependent longitudinal density-density
correlation function goes into the well known classical result.
The computational results for graphene-graphene and graphene-Au
plate Casimir interactions obtained using the
temperature-dependent
density-density correlation functions found by us are exactly
coinciding with those obtained using the temperature-dependent
polarization tensor.

One can conclude that an equivalence of the two approaches to
calculation of the van der Waals and Casimir forces in layered
systems including graphene demonstrated in this paper provides
a reliable foundation for the comparison between experiment
and theory.


\end{document}